\documentclass[sigconf]{acmart}

\copyrightyear{2019}
\acmYear{2019}
\setcopyright{iw3c2w3}
\acmConference[WWW '19]{Proceedings of the 2019 World Wide Web Conference}{May 13--17, 2019}{San Francisco, CA, USA}
\acmBooktitle{Proceedings of the 2019 World Wide Web Conference (WWW '19), May 13--17, 2019, San Francisco, CA, USA}
\acmPrice{}
\acmDOI{10.1145/3308558.3313614}
\acmISBN{978-1-4503-6674-8/19/05}
\usepackage{natbib}
\usepackage{url}
\usepackage{booktabs} % For formal tables
\usepackage{algorithm}
\usepackage{algorithmic}
\usepackage{amssymb}
\usepackage{array}
\usepackage[normalem]{ulem}
\usepackage{footmisc}
\usepackage{graphicx}
\usepackage{epstopdf}
\usepackage{multirow}
\usepackage{enumitem}
\usepackage[skip=0pt]{caption}
\usepackage{subfigure}
\usepackage{xcolor}
\usepackage{pifont}
\usepackage{bm}
\usepackage{acronym}
\acrodef{CNN}{Convolutional Neural Network}
\acrodef{DCNN}{DeConvolutional Neural Network}
\acrodef{RNN}{Recurrent Neural Network}
\acrodef{Bi-LSTM}{Bidirectional Long Short-Term Memory Network}
\acrodef{BPR}{Bayesian Personalized Ranking}
\acrodef{ELBO}{Evidence Lower BOund}
\acrodef{AUC}{Area Under the ROC Curve}
\acrodef{MRR}{Mean Reciprocal Rank}
\acrodef{VU}{visual understanding}
\acrodef{VM}{visual matching}
\acrodef{SIFT}{Scale Invariant Feature Transform}
\acrodef{HOG}{Histograms of Oriented Gradient}
\acrodef{LBP}{Local Binary Pattern}
\acrodef{SVM}{Support Vector Machine}
\acrodef{GM}{Graphical Models}
\acrodef{CRF}{Conditional Random Field}
\acrodef{CTM}{Correlated Topic Models}
\acrodef{VAE}{Variational Auto-Encoder}
\acrodef{FARM}{FAshion Recommendation Machine}
\newcount\colveccount
\newcommand*\colvec[1]{
        \global\colveccount#1
        \begin{matrix}
        \colvecnext
}
\def\colvecnext#1{
        #1
        \global\advance\colveccount-1
        \ifnum\colveccount>0
                \\
                \expandafter\colvecnext
        \else
                \end{matrix}
        \fi
}

\begin{document}
\title{Improving Outfit Recommendation with Co-supervision of Fashion Generation}

\author{Yujie Lin}
\authornote{Co-first author.}
\orcid{}
\affiliation{%
\institution{Shandong University}
\city{Qingdao}
\country{China}
}
\email{yu.jie.lin@outlook.com}

\author{Pengjie Ren}
\authornotemark[1]
\orcid{}
\affiliation{%
\institution{University of Amsterdam}
\city{Amsterdam}
\country{The Netherlands}
}
\email{p.ren@uva.nl}

\author{Zhumin Chen}
\orcid{}
\affiliation{%
\institution{Shandong University}
\city{Qingdao}
\country{China}
}
\email{chenzhumin@sdu.edu.cn}

\author{Zhaochun Ren}
\orcid{}
\affiliation{%
\institution{Shandong University}
\city{Qingdao}
\country{China}
}
\email{zhaochun.ren@sdu.edu.cn}

\author{Jun Ma}
\orcid{}
\affiliation{%
\institution{Shandong University}
\city{Qingdao}
\country{China}
}
\email{majun@sdu.edu.cn}

\author{Maarten de Rijke}
\orcid{0000-0002-1086-0202}
\affiliation{%
\institution{University of Amsterdam}
\city{Amsterdam}
\country{The Netherlands}
}
\email{derijke@uva.nl}

\begin{abstract}
The task of fashion recommendation includes two main challenges: \emph{visual understanding} and \emph{visual matching}.
Visual understanding aims to extract effective visual features.
Visual matching aims to model a human notion of compatibility to compute a match between fashion items.
Most previous studies rely on recommendation loss alone to guide visual understanding and matching.
Although the features captured by these methods describe basic characteristics (e.g., color, texture, shape) of the input items, they are not directly related to the visual signals of the output items (to be recommended).
This is problematic because the aesthetic characteristics (e.g., style, design), based on which we can directly infer the output items, are lacking.
Features are learned under the recommendation loss alone, where the supervision signal is simply whether the given two items are matched or not.

To address this problem, we propose a neural co-supervision learning framework, called the \ac{FARM}.
\ac{FARM} improves \acl{VU} by incorporating the supervision of generation loss, which we hypothesize to be able to better encode aesthetic information.
\ac{FARM} enhances \acl{VM} by introducing a novel layer-to-layer matching mechanism to fuse aesthetic information more effectively, and meanwhile avoiding paying too much attention to the generation quality and ignoring the recommendation performance.

Extensive experiments on two publicly available datasets show that \ac{FARM} outperforms state-of-the-art models on outfit recommendation, in terms of AUC and MRR.
Detailed analyses of generated and recommended items demonstrate that \ac{FARM} can encode better features and generate high quality images as references to improve recommendation performance.
\end{abstract}
%
% The code below should be generated by the tool at
% http://dl.acm.org/ccs.cfm
% Please copy and paste the code instead of the example below.
%
%Please go to: http://dl.acm.org/ccs_flat.cfm and choose as many classifiers as you feel appropriate to your submission and the level of relevance: high, medium, low. Then click "[continue]"to choose additional classifiers. When you are done choosing your choices of classifiers, click on the link near the top of the pop-up window "[View CCS TeX Code]"



\maketitle

% !TEX root = ./www2019-fp-yujie-pengjie.tex

\section{Introduction}

Fashion recommendation has attracted increasing attention~\cite{Hu2015,Jaradat:2017:DCF:3109859.3109861,DBLP:conf/icdm/KangFWM17} for its potentially wide applications in fashion-oriented online communities such as, e.g., Polyvore\footnote{\url{http://www.polyvore.com/}} and Chictopia.\footnote{\url{http://www.chictopia.com/}}
By recommending fashionable items that people may be interested in, fashion recommendation can promote the development of online retail by stimulating people's interests and participation in online shopping.
In this paper, we target \emph{outfit recommendation}, that is, given a top (i.e., upper garment), we need to recommend a list of bottoms (e.g., trousers or skirts) from a large collection that best match the top, and vice versa. 
Specifically, we allow users to provide some descriptions as conditions that the recommended items should accord with as much as possible. 

Unlike conventional recommendation tasks, outfit recommendation faces two main challenges: \emph{visual understanding} and \emph{visual matching}.
Visual understanding aims to extract effective features by building a deep understanding of fashion item images. Visual matching requires modeling a human notion of the compatibility between fashion items \cite{Song2017}, which involves matching features such as color and shape etc.
Early studies into outfit recommendation rely on feature engineering for \acl{VU} and traditional machine learning  for \acl{VM} \cite{Jagadeesh2014}.
For example, \citet{Iwata2011} define three types of feature, i.e., color, texture and local descriptors such as \ac{SIFT} (for \acl{VU}), and propose a recommendation model based on \ac{GM} (for \acl{VM}).
\citet{Liu2012} define five types of feature including \ac{HOG} \cite{Dalal:2005:HOG:1068507.1069007}, \ac{LBP} \cite{Ahonen:2006:FDL:1175897.1176245}, color moment, color histogram and skin descriptor \cite{DBLP:conf/iccv/BourdevMM11} (for \acl{VU}), and propose a latent \ac{SVM} based recommendation model (for \acl{VM}).

Recently, neural networks have been applied to address the challenges of fashion recommendation: \citet{Song2017} use a pre-trained \ac{CNN} (on ImageNet) to extract visual features (for \acl{VU}).
Then, they employ a separate \ac{BPR} \cite{Rendle2009BPR} method to exploit pairwise preferences between tops and bottoms (for \acl{VM}).
\citet{2018arXiv180608977L} propose to train feature extraction (for \acl{VU}) and preference prediction (for \acl{VM}) in a single back-propagation scheme.
They introduce a mutual attention mechanism into \ac{CNN} to improve feature extraction.
The visual features captured by these methods only describe basic characteristics (e.g., color, texture, shape) of the input items, which lack aesthetic characteristics (e.g., style, design) to describe the output items (to be recommended).
Visual understanding and matching are conducted based on recommendation loss alone, where the supervision signal is just whether two given items are matched or not and no supervision is available to directly connect the visual signals of the fashion items.
Recently, some studies have realized the importance of modeling aesthetic information.
For example, \citet{DBLP:conf/aaai/MaJZFLT17} build a universal taxonomy to quantitatively describe aesthetic characteristics of clothing.
\citet{Yu:2018:ACR:3178876.3186146} propose to encode aesthetic information by pre-training models on aesthetic assessment datasets.
However, none of them is for outfit recommendation and none improves visual understanding and matching like we do.

In this paper, we address the challenges of outfit recommendation from a novel perspective by proposing a neural co-supervision learning framework, called \acfi{FARM}.
\ac{FARM} enhances \acl{VU} and \acl{VM} with the \emph{joint supervision} of generation and recommendation learning. 
Let us explain.
By incorporating the generation process as a supervision signal, \ac{FARM} is able to encode more aesthetic characteristics, based on which we can directly generate the output items.
\ac{FARM} enhances \acl{VM} by incorporating a novel layer-to-layer matching mechanism to evaluate the matching score of generated and candidate items at different neural layers; in this manner \ac{FARM} fuses the generation features from different visual levels to improve the recommendation performance.
This layer-to-layer matching mechanism also ensures that \ac{FARM} avoids paying too much attention to the generation quality and ignoring the recommendation performance.
To the best of our knowledge, \ac{FARM} is the first end-to-end learning framework that improves outfit recommendation with joint modeling of fashion generation.

Extensive experimental results conducted on two publicly available datasets show that \ac{FARM} outperforms  state-of-the-art models on outfit recommendation, in terms of AUC and MRR.
To further demonstrate the advantages of \ac{FARM}, we conduct several analyses and case studies.

To sum up, our contributions can be summarized as follows:

\begin{itemize}[nosep,leftmargin=*]
\item We propose a neural co-supervision learning framework, \ac{FARM}, for outfit recommendation that simultaneously yields recommendation and generation.
\item We propose a layer-to-layer matching mechanism that acts as a bridge between generation and recommendation, and improves recommendation by leveraging generation features.
\item Our proposed approach is shown to be effective in experiments on two large-scale datasets.
\end{itemize}

% !TEX root = ./www2019-fp-yujie-pengjie.tex
\section{Related Work}
\begin{figure*}
 \centering
 \subfigure{
 \includegraphics[width=1.0\textwidth]{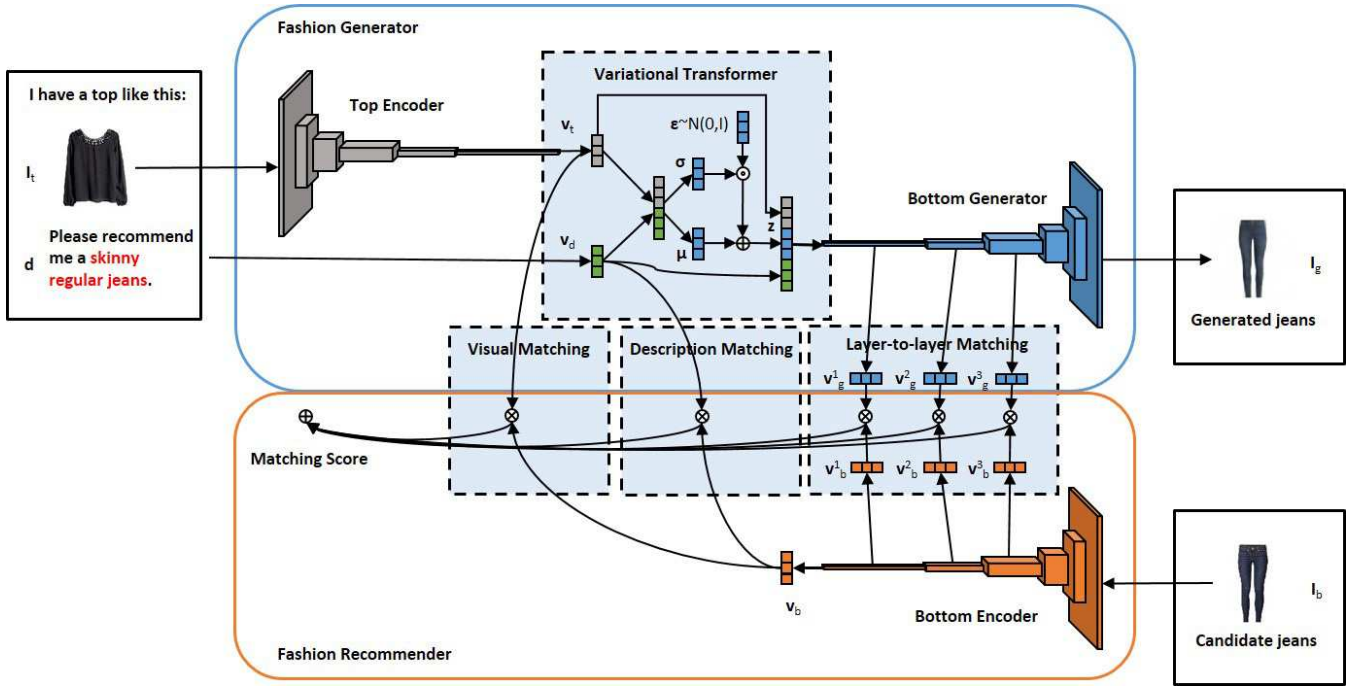}}
 \caption{Overview of \ac{FARM}. The fashion generator (top) uses a variational transformer to learn a special Gaussian distribution for a given top image $\textbf{I}_t$ and a given bottom description $\textbf{d}$. It then generates a bottom image $\textbf{I}_g$ to match $\textbf{I}_t$ and $\textbf{d}$. The fashion recommender (bottom) evaluates the matching score between the recommended bottom image $\textbf{I}_b$ and $(\textbf{I}_t,\textbf{d})$ pair from three angles, i.e., visual matching, description matching, and layer-to-layer matching.}
  \label{f_3_1}
\end{figure*}

We survey related work on fashion recommendation by focusing on the two main challenges in the area: \acl{VU} and \acl{VM}.

\subsection{Visual understanding}

One branch of studies aims at extracting better features to improve the \acl{VU} of fashion items.

For instance, \citet{Iwata2011} propose a recommender system for clothing coordinates using full-body photographs from fashion magazines.
They extract visual features, such as color, texture and local descriptors such as \ac{SIFT}, and use a probabilistic topic model for \acl{VU} of coordinates from these features.
\citet{Liu2012} target occasion-oriented clothing recommendation. 
Given a user-input event, e.g., wedding, shopping or dating, their model recommends the most suitable clothing from the user's own clothing photo album.
They adopt clothing attributes (e.g., clothing category, color, pattern) for better \acl{VU}.
\citet{Jagadeesh2014} describe a visual recommendation system for street fashion images.
They mainly focus on color modeling in terms of \acl{VU}.

The studies listed above achieve \acl{VU} mostly based on feature engineering and conventional machine learning techniques.
With the availability of large scale fashion recommendation datasets and the rapid development of deep learning models, several recent publications turn to neural networks for fashion recommendation.
\acp{CNN} are certainly widely employed \cite{McAuley2015,Li2017MiningFO}.
\citet{DBLP:conf/aaai/MaJZFLT17} build a taxonomy based on a theory of aesthetics to describe aesthetic features of fashion items quantitatively and universally. 
Then they capture the internal correlation in clothing collocations by a novel fashion-oriented multi-modal deep learning based model.
\citet{Song2017} use a pre-trained \ac{CNN} on ImageNet to extract visual features.
Then, to improve \acl{VU} with contextual information (such as titles and categories), they propose to use multi-modal auto-encoders to exploit the latent compatibility of visual and contextual features.
\citet{Han2017} enrich \acl{VU} by incorporating sequential information by using a \ac{Bi-LSTM} to predict the next item conditioned on previous ones. 
They further inject attribute and category information as a kind of regularization to learn a visual-semantic space by regressing visual features to their semantic representations. 
\citet{DBLP:conf/icdm/KangFWM17} use a \ac{CNN}-F~\cite{Chatfield14} to learn image representations and train a personalized fashion recommendation system jointly.
Besides, they devise a personalized fashion design system based on the learned \ac{CNN}-F and user representations.
\citet{Yu:2018:ACR:3178876.3186146} propose to introduce aesthetic information into fashion recommendation. 
To achieve this, they extract aesthetic features using a pre-trained  brain-inspired deep structure on the aesthetic assessment task.
\citet{2018arXiv180608977L} enhance \acl{VU} by jointly modeling fashion recommendation and user comment generation, where the visual features learned with a \ac{CNN} are enriched because they are related to the generation of user comments.

Even though there is a growing number of studies on better \acl{VU} for fashion recommendation, none of them takes fashion generation into account like we do in this paper.

\subsection{Visual matching}

Early studies into \acl{VM} are based on conventional machine learning methods. 
\citet{Iwata2011} use a topic model to learn the relation between photographs and recommend a bottom that has the closest topic proportions to those of the given top. 
\citet{Liu2012} employ an \ac{SVM} for recommendation, which has a term describing the relationship between visual features and attributes of tops and bottoms. 
\citet{7298688} predict the popularity of an outfit to implicitly learn its compatibility by a \ac{CRF} model. 
\citet{McAuley2015} measure the compatibility between clothes by learning a distance metric with pre-trained \ac{CNN} features. 
\citet{Hu2015} propose a functional pairwise interaction tensor factorization method to model the interactions between fashion items of different categories. 
\citet{Hsiao2017Creating} develop a submodular objective function to capture the key ingredients of visual compatibility in outfits. 
They propose a topic model namely \ac{CTM} to generate compatible outfits learned from unlabeled images of people wearing outfits.

Recently, deep learning methods have been used widely in the fashion recommendation community. 
\citet{Veit2015Learning} train an end-to-end Siamese \ac{CNN} network to learn a feature transformation from images to a latent compatibility space. 
\citet{Oramas2016Modeling} mine mid-level elements from \acp{CNN} to model the compatibility of clothes.
\citet{Li2017MiningFO} use a \ac{RNN} to predict whether an outfit is popular, which also implicitly learns the compatibility relation between fashion items. 
\citet{Han2017} further train a \ac{Bi-LSTM} to sequentially predict the next item conditioned on the previous ones for learning their compatibility relationship. 
\citet{Song2017} employ a dual auto-encoder network to learn the latent compatibility space where they use the \ac{BPR} model to jointly model the relation between visual and contextual modalities and implicit preferences among fashion items. 
\citet{Song2018Neural} consider the knowledge about clothing matching and follow a teacher-student scheme to encode the fashion domain knowledge in a traditional neural network. 
And they  introduce an attentive scheme to the knowledge distillation procedure to flexibly assign rule confidence. 
\citet{Nakamura2018Outfit} present an architecture containing three subnetworks, i.e., VSE (Visual-Semantic Embedding), \ac{Bi-LSTM} and SE (Style Embedding) modules, to model the matching relation between different items to generate outfits. 
\citet{2018arXiv180608977L} propose a mutual attention mechanism into \acp{CNN} to model the compatibility between different parts of images of fashion items. 

Although there are many studies on improving \acl{VM}, none of them considers connecting it with fashion generation.

% !TEX root = ./www2019-fp-yujie-pengjie.tex

\section{Neural Fashion Recommendation}
\begin{figure*}
 \centering
 \subfigure[Encoder]{
 \label{f_3_2a}
 \includegraphics[width=0.8\textwidth]{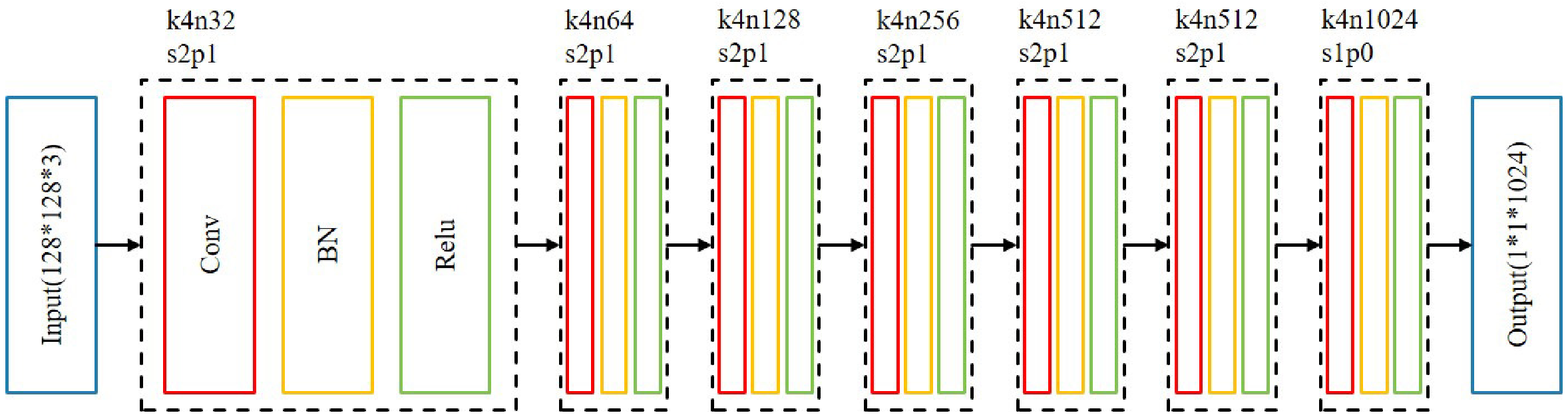}}
 \subfigure[Generator]{
 \label{f_3_2b}
 \includegraphics[width=1.0\textwidth]{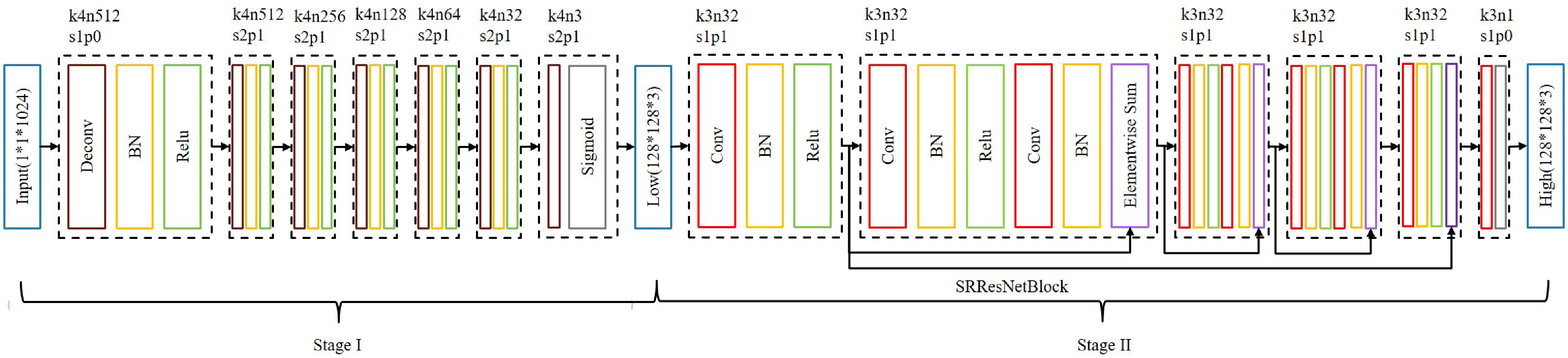}}
 \caption{Details of the encoder and the generator in \ac{FARM}, where $k$ represents kernel size, $n$ represents the number of channels, $s$ represents strides and $p$ represents padding.}
 \label{f_3_2}
\end{figure*}

\subsection{Overview}
Given a top $t$ from a pool $\mathcal{T} = \{t_1, t_2, \ldots , t_{N_t}\}$ and a user's description $d$ for the target bottom, the \emph{bottom recommendation task} is to recommend a list of bottoms from a candidate pool $\mathcal{B} = \{b_1, b_2, \ldots , b_{N_b}\}$.
Similarly, the \emph{top recommendation task} is to recommend a ranked list of tops for a given bottom and top description pair.
Here, we use bottom recommendation as the setup to introduce our framework \ac{FARM}.

As shown in Figure~\ref{f_3_1}, \ac{FARM} consists of two parts, i.e., a fashion generator (for \acl{VU}) and a fashion recommender (for \acl{VM}), where the fashion generator is actually an auxiliary module for recommendation.
For the fashion generator, we use a \ac{CNN} as the top encoder to extract the visual features from a given top image $\textbf{I}_t$.
We learn the semantic representation for the bag-of-words vector $\textbf{d}$ of a given bottom description.
Then we use a variational transformer to learn the mapping from the bottom distribution to a specific Gaussian distribution that is based on the visual features of $\textbf{I}_t$ and the semantic representation of $\textbf{d}$.
Finally, we sample a random vector from the Gaussian distribution and input it to a \ac{DCNN}~\cite{Zeiler2011Adaptive} (as bottom generator) to generate a bottom image $\textbf{I}_g$ that matches $\textbf{I}_t$ and $\textbf{d}$, which explicitly forces the top encoder to encode more aesthetic matching information into the visual features.
For the fashion recommender, we also employ a \ac{CNN} as the bottom encoder to extract the visual features from a candidate bottom image $\textbf{I}_b$.
Then we evaluate the matching score between $\textbf{I}_b$ and $(\textbf{I}_t, \textbf{d})$ pair from three angles, namely the visual matching between $\textbf{I}_b$ and $\textbf{I}_t$, the description matching between $\textbf{I}_b$ and $\textbf{d}$, and the layer-to-layer matching between $\textbf{I}_b$ and $\textbf{I}_g$ which leverages the generation information to improve the recommendation.
\ac{FARM} jointly trains the fashion generator and fashion recommender.
Next we will detail each of these two main parts.

\subsection{Fashion generator}
Given an image $\textbf{I}_t$ of a top $t$ and the bag-of-words vector $\textbf{d}$ of a bottom description $d$, the fashion generator needs to generate a bottom image $\textbf{I}_g$ that not only matches $\textbf{I}_t$, but also meets $\textbf{d}$ as much as possible.
We enforce the extracted visual features from $\textbf{I}_t$ to contain the information about its matching bottom by using the generator as a supervision signal.
The generated image can be seen as a reference for recommendation.

Specifically, for a generated bottom image $\textbf{I}_g$ that matches $\textbf{I}_t$ and $\textbf{d}$, the aim of the fashion generator is to maximize Eq.~\ref{p(I_g|I_t,d)}:
\begin{equation}
\label{p(I_g|I_t,d)}
p(\textbf{I}_g|\textbf{I}_t,\textbf{d}) = \int_{\textbf{z}}p(\textbf{I}_g|\textbf{z},\textbf{I}_t,\textbf{d})p(\textbf{z}|\textbf{I}_t,\textbf{d})\mathrm{d}\textbf{z},
\end{equation}
where $p(\textbf{z}|\textbf{I}_t,\textbf{d})$ is the top encoder, $p(\textbf{I}_g|\textbf{z},\textbf{I}_t,\textbf{d})$ is the bottom generator, and $\textbf{z}$ is the latent variable.
Because the integral of the marginal likelihood shown in Eq.~\ref{p(I_g|I_t,d)} is intractable, inspired by variational inference~\cite{Blei2017Variational}, we first find the \ac{ELBO} of $p(\textbf{I}_g|\textbf{I}_t,\textbf{d})$, as shown in Eq.~\ref{ELBO}:
\begin{eqnarray}
\label{ELBO}
\mathrm{ELBO} = \mathbb{E}_{\textbf{z}\sim{q(\textbf{z}|\textbf{I}_t,\textbf{d})}}[\log{p(\textbf{I}_g|\textbf{z},\textbf{I}_t,\textbf{d})}]\nonumber\\
-\mathrm{KL}[q(\textbf{z}|\textbf{I}_t,\textbf{d}){\parallel}p(\textbf{z}|\textbf{I}_t,\textbf{d})],
\end{eqnarray}
where $q(\textbf{z}|\textbf{I}_t,\textbf{d})$ is the approximation of the intractable true posterior $p(\textbf{z}|\textbf{I}_g,\textbf{I}_t,\textbf{d})$.
The following inequality holds for the \ac{ELBO}:
\begin{equation}
\log{p(\textbf{I}_g|\textbf{I}_t,\textbf{d})} \geqslant \mathrm{ELBO}.
\end{equation}
Hence, we can maximize the \ac{ELBO} so as to maximize $\log{p(\textbf{I}_g|\textbf{I}_t,\textbf{d})}$.
The \ac{ELBO} contains three components:  $q(\textbf{z}|\textbf{I}_t,\textbf{d})$, $p(\textbf{z}|\textbf{I}_t,\textbf{d})$ and $p(\textbf{I}_g|\textbf{z},\textbf{I}_t,\textbf{d})$.
Below we explain each component in detail.

\subsubsection{$q(\textbf{z}|\textbf{I}_t,\textbf{d})$ and $p(\textbf{z}|\textbf{I}_t,\textbf{d})$.}
We propose a variational transformer (as shown in Figure~\ref{f_3_1}) to model these two components, which transforms $\textbf{I}_t,\textbf{d}$ into a latent variable $\textbf{z}$.
As with previous work~\cite{Kingma2014Auto, Rezende2014Stochastic}, we assume that $q(\textbf{z}|\textbf{I}_t,\textbf{d})$ and $p(\textbf{z}|\textbf{I}_t,\textbf{d})$ are Gaussian distributions, i.e.,
\begin{equation}
q(\textbf{z}|\textbf{I}_t,\textbf{d}) \sim \mathcal{N}(\textbf{z};\bm{\mu},\bm{\sigma}^2),\quad
p(\textbf{z}|\textbf{I}_t,\textbf{d}) \sim \mathcal{N}(0,1),
\end{equation}
where $\bm{\mu}$ and $\bm{\sigma}$ denote the variational mean and standard deviation respectively, which are calculated with our top encoder and variational transformer as follows.

Specifically, for a top image $\textbf{I}_t$ of size $128\times{128}$ with $3$ channels, we first use a \ac{CNN}, i.e., \emph{the top encoder} (as shown in Figure~\ref{f_3_2a}) to extract visual features $\textbf{F}_t$:
\begin{equation}
\textbf{F}_t = \textbf{CNN}(\textbf{I}_t),
\end{equation}
where $\textbf{F}_t\in\mathbb{R}^{W\times{H}\times{D}}$, $W$ and $H$ are the width and height of the output feature maps, respectively, and $D$ is the number of output feature maps. 
And we flatten $\textbf{F}_t$ into a vector $\textbf{f}_t\in\mathbb{R}^N$, where $N = W\times{H}\times{D}$, and project $\textbf{f}_t$ to the visual representation $\textbf{v}_t$:
\begin{equation}
\textbf{v}_t = \mathrm{sigmoid}(\textbf{W}_{vt}\textbf{f}_t+\textbf{b}_{vt}),
\end{equation}
where $\textbf{W}_{vt}\in\mathbb{R}^{e\times{N}}$, $\textbf{v}_t$ and $\textbf{b}_{vt}\in\mathbb{R}^e$, and $e$ is the size of the representation. 

Besides the top image, \ac{FARM} also allows users to give a natural language description $\textbf{d}$, which describes the ideal bottom they want.
In order to take into account the description $\textbf{d}$, we follow Eq.~\ref{v_d} to get the semantic representation $\textbf{v}_d$:
\begin{equation}
\label{v_d}
\textbf{v}_d = \mathrm{sigmoid}(\textbf{W}_d\textbf{d}),
\end{equation}
where $\textbf{v}_d\in\mathbb{R}^e$, $\textbf{d}\in\mathbb{R}^{D_d}$, $D_d$ is the vocabulary size, and $\textbf{W}_d\in\mathbb{R}^{e\times{D_d}}$ is the visual semantic word embedding matrix~\cite{Nakamura2018Outfit}, which transforms words from the textual space to the visual space.
Specially, when $d$ is an empty description, $\textbf{v}_d$ is a zero vector.

Then \emph{the variational transformer} uses the visual representation $\textbf{v}_t$ and the semantic representation $\textbf{v}_d$ to calculate the mean $\bm{\mu}$ and standard deviation $\bm{\sigma}$ for $q(\textbf{z}|\textbf{I}_t,\textbf{d})$:
\begin{equation}
\begin{split}
\bm{\mu} = {} & \textbf{W}_{\mu{t}}\textbf{v}_t+\textbf{W}_{\mu{d}}\textbf{v}_d+\textbf{b}_{\mu} \\
\log{\bm{\sigma}^2} = {} & \textbf{W}_{\sigma{t}}\textbf{v}_t+\textbf{W}_{\sigma{d}}\textbf{v}_d+\textbf{b}_{\sigma},
\end{split}
\end{equation}
where $\textbf{W}_{\mu{t}}$, $\textbf{W}_{\mu{d}}$, $\textbf{W}_{\sigma{t}}$ and $\textbf{W}_{\sigma{d}}\in\mathbb{R}^{k\times{e}}$, $\bm{\mu}$, $\bm{\sigma}$, $\textbf{b}_{\mu}$ and $\textbf{b}_{\sigma}\in\mathbb{R}^k$, and $k$ is the size of latent variable $\textbf{z}$.
The latent variable $\textbf{z}$ can be calculated by the reparameterization trick~\cite{Kingma2014Auto, Rezende2014Stochastic}:
\begin{equation}
\bm{\epsilon}\sim\mathcal{N}(0,1),\quad 
\textbf{z} = \bm{\mu}+\bm{\sigma}\otimes\bm{\epsilon},
\end{equation}
where $\bm{\epsilon}$ and $\textbf{z}\in\mathbb{R}^k$, and $\bm{\epsilon}$ is the auxiliary noise variable.
By the reparameterization trick, we make sure $\textbf{z}$ is a random vector sampled from $\mathcal{N}(\textbf{z};\bm{\mu},\bm{\sigma}^2)$.

\subsubsection{$p(\textbf{I}_g|\textbf{z},\textbf{I}_t,\textbf{d})$.}
We use the bottom generator (as shown in Figure~\ref{f_3_2b}) to generate $\textbf{I}_g$ from the variable $\textbf{z}$.
We also assume $p(\textbf{I}_g|\textbf{z},\textbf{I}_t,\textbf{d})$ is a Gaussian distribution~\cite{Kingma2014Auto, Rezende2014Stochastic}, i.e.,
\begin{equation}
p(\textbf{I}_g|\textbf{z},\textbf{I}_t,\textbf{d}) \sim \mathcal{N}(g(\textbf{z},\textbf{I}_t,\textbf{d}),\bm{\sigma}^2),
\end{equation}
where $g$ is the bottom generator.
% and we \todo{HUH: hope} $g(\textbf{z},\textbf{I}_t,\textbf{d})=\textbf{I}_g^*$.

Specifically, we first follow Eq.~\ref{f_g} to obtain the basic visual feature vector $\textbf{f}_g$:
\begin{equation}
\label{f_g}
\textbf{f}_g = \mathrm{relu}(\textbf{W}_{gz}\textbf{z}+\textbf{W}_{gt}\textbf{v}_t+\textbf{W}_{gd}\textbf{v}_d+\textbf{b}_g),
\end{equation}
where $\textbf{f}_g$ and $\textbf{b}_g\in\mathbb{R}^N$, $\textbf{W}_{gz}\mathbb{R}^{N\times{k}}$, $\textbf{W}_{gt}$ and $\textbf{W}_{gd}\in\mathbb{R}^{N\times{e}}$.
Then we reshape $\textbf{f}_g$ into a $3$-D tensor $\textbf{F}_g\in\mathbb{R}^{W\times{H}\times{D}}$, which is the reverse operation to what we do for $\textbf{F}_t$.
Finally, we use a \ac{DCNN}, i.e., \emph{the bottom generator} to generate the bottom image $\textbf{I}_g$:
\begin{equation}
\textbf{I}_g = \textbf{DCNN}(\textbf{F}_g),
\end{equation}
where $\textbf{I}_g\in\mathbb{R}^{128\times{128}\times{3}}$. To avoid generating blurry images~\cite{Bao2017CVAE}, we divide the process of image generation into two stages~\cite{8237891, DBLP:journals/corr/CaiGJ17}.
The first stage is an ordinary deconvolutional neural network that generates low-resolution images.
The second stage is similar to the super-resolution residual network (SRResNet)~\cite{Ledig2017Photo}, which accepts the images from the first stage and refines them to generate high quality ones.
The \ac{DCNN} is meant to capture high-level aesthetic features of the bottoms to be recommended~\citep{Zeiler2011Adaptive,Zeiler2014Visualizing}.
Besides, in order to generate the bottom, the generation process also forces the top encoder to capture more aesthetic information.

During training, we first sample a $z$ from $q(\textbf{z}|\textbf{I}_t,\textbf{d})$.
Then we generate $\textbf{I}_g$ with $g(\textbf{z},\textbf{I}_t,\textbf{d})$. 
During testing, in order to avoid the randomness introduced by $\bm{\epsilon}$, we directly generate $\textbf{I}_g$ by $g(\textbf{z}=\bm{\mu},\textbf{I}_t,\textbf{d})$.

\subsection{Fashion recommender}
Given the image $\textbf{I}_b$ of a bottom $b$, the fashion recommender needs to evaluate the matching score between $\textbf{I}_b$ and the pair $(\textbf{I}_t,\textbf{d})$.
Specifically, we first use the \emph{bottom encoder} (as shown in Figure~\ref{f_3_2a}), which has the same structure as the top encoder (parameters not shared), to extract visual features $\textbf{F}_b\in\mathbb{R}^{W\times{H}\times{D}}$ from $\textbf{I}_b$.
Then we flatten $\textbf{F}_b$ into a vector $\textbf{f}_b\in\mathbb{R}^N$ and project $\textbf{f}_b$ to the visual representation $\textbf{v}_b$.
Next, we calculate the matching score between $\textbf{I}_b$ and the pair $(\textbf{I}_t,\textbf{d})$ in three ways.

\subsubsection{Visual matching}
We propose visual matching to evaluate the compatibility between $\textbf{I}_b$ and $\textbf{I}_t$ based on their visual features.
Specifically, we calculate the visual matching score $s_v$ between $\textbf{I}_b$ and $\textbf{I}_t$ by Eq.~\ref{s_v}:
\begin{equation}
\label{s_v}
s_v = \textbf{v}_b^T\textbf{v}_t.
\end{equation}

\subsubsection{Description matching}
For evaluating the matching degree between $\textbf{I}_b$ and $\textbf{d}$, we propose to match descriptions.
The description matching score $s_d$ between $\textbf{I}_b$ and $\textbf{d}$ is calculated by Eq.~\ref{s_d}:
\begin{equation}
\label{s_d}
s_d = \textbf{v}_b^T\textbf{v}_d.
\end{equation}
Note that if $d$ does not contain any word, $s_d$ equals 0.

\subsubsection{Layer-to-layer matching}
As we will demonstrate in our experiments in Section~\ref{section:pma}, a simple combination of generation and recommendation is not able to improve the recommendation performance.
The reason is that there is no direct connection between generation and recommendation, which results in two issues.
First, the aesthetic information from the generation process cannot be used effectively.
Second, the generation process might introduce features that are only helpful for generation while unhelpful for recommendation.
To overcome these issues, we propose a layer-to-layer matching mechanism.
Specifically, we denote the visual features of the $l$-th \ac{CNN} layer in the bottom encoder as $\textbf{F}^l_b\in\mathbb{R}^{W^l\times{H^l}\times{D^l}}$.
And we denote the visual features of the corresponding \ac{DCNN} layer, which has the same size as $\textbf{F}^l_b$, in the bottom generator as $\textbf{F}^l_g\in\mathbb{R}^{W^l\times{H^l}\times{D^l}}$.
Then, we reshape $\textbf{F}^l_b = [\textbf{f}^l_{b,1}, \ldots, \textbf{f}^l_{b,S}]$ by flattening the width and height of the original $\textbf{F}^l_b$, where $S=W^l\times{H^l}$ and $\textbf{f}^l_{b,i}\in\mathbb{R}^{D^l}$.
And we can consider $\textbf{f}^l_{b,i}$ as the visual features of the $i$-th location of $\textbf{I}_b$. 
We perform global-average-pooling in $\textbf{F}^l_b$ to get the global visual features $\textbf{f}^l_b\in\mathbb{R}^{D^l}$:
\begin{equation}
\textbf{f}^l_b = \frac{1}{S}\sum^S_{i=1}\textbf{f}^l_{b,i}.
\end{equation}
We project $\textbf{f}^l_b$ to the visual representation $\textbf{v}^l_b\in\mathbb{R}^e$:
\begin{equation}
\textbf{v}^l_b = \mathrm{sigmoid}(\textbf{W}^l_{vb}\textbf{f}^l_b+\textbf{b}^l_{vb}),
\end{equation}
where $\textbf{W}^l_{vb}\in\mathbb{R}^{e\times{D^l}}$ and $\textbf{b}^l_{vb}\in\mathbb{R}^e$.
The same operations apply to $\textbf{F}^l_g$ to get $\textbf{v}^l_g$.
Then we calculate the dot product between $\textbf{v}^l_b$ and $\textbf{v}^l_g$, which represents the matching degree $s^l_g$ between $\textbf{I}_b$ and $\textbf{I}_g$ in the $l$-th visual level:
\begin{equation}
\label{s^l_g}
s^l_g = {\textbf{v}^l_b}^T\textbf{v}^l_g.
\end{equation}
For different visual levels, we sum all $s^l_g$ to get the matching score $s_g$ between $\textbf{I}_b$ and $\textbf{I}_g$:
\begin{equation}
s_g = \sum_{l\in{L}}s^l_g,
\end{equation}
where $L$ is the selected \ac{CNN} layer set for layer-to-layer matching.

Finally, the total matching score $s$ between $\textbf{I}_b$ and the pair $(\textbf{I}_t,\textbf{d})$ is defined as follows:
\begin{equation}
\label{ms}
s = s_v+s_d+s_g.
\end{equation}

\subsection{Co-supervision learning framework}
For \ac{FARM}, we train the fashion generator and the fashion recommender jointly with a co-supervision learning framework.

Specifically, for the generation part, we regard the image $\textbf{I}_p$ of a positive bottom $p$, which not only matches the given top $\textbf{I}_t$ but also meets the given description $\textbf{d}$, as the generation target.
And we denote the generated bottom image in the first stage as $\textbf{I}^1_g$, and denote the generated bottom image in the second stage as $\textbf{I}^2_g$.
Then, the first loss is to maximize the first term in \ac{ELBO}, which is Eq.~\ref{L_{gen}}:
\begin{equation}
\label{L_{gen}}
\mathcal{L}_{\mathit{gen}}(t,d,p) = \frac{1}{2}\|\textbf{I}^1_g-\textbf{I}_p\|^2_2+\|\textbf{I}^2_g-\textbf{I}_p\|.
\end{equation}
The second loss is to minimize the second term in \ac{ELBO}, which is Eq.~\ref{L_{kl}}:
\begin{equation}
\label{L_{kl}}
\mathcal{L}_{\mathit{kl}}(t,d,p) = \frac{1}{2}\sum^k_{i=1}(1+\log\bm{\sigma}^2_i-\bm{\mu}^2_i-\bm{\sigma}^2_i),
\end{equation}
where $\bm{\mu}_i$ and $\bm{\sigma}_i$ are the $i$-th elements in $\bm{\mu}$ and $\bm{\sigma}$ respectively.

For the recommendation part, we employ \ac{BPR}~\cite{Rendle2009BPR} as the loss:
\begin{equation}
\label{L_{bpr}}
\mathcal{L}_{\mathit{bpr}}(t,d,p,n) = -\log(\mathrm{sigmoid}(s_p-s_n)),
\end{equation}
where $s_p$ and $s_n$ are the matching scores of a positive bottom $\textbf{I}_p$ and a negative bottom $\textbf{I}_n$, respectively (calculated with Eq.~\ref{ms}).
$\textbf{I}_n$ (image of bottom $n$) is randomly sampled.

The total loss function can be defined as follows:
\begin{equation}
\mathcal{L} = \sum_{(t,d,p,n)\in\mathcal{D}}\mathcal{L}_{\mathit{gen}}(t,d,p)+\mathcal{L}_{\mathit{kl}}(t,d,p)+\mathcal{L}_{\mathit{bpr}}(t,d,p,n),
\end{equation}
where $\mathcal{D}=\{(t,d,p,n)|t\in\mathcal{T}, d\in\mathcal{D}_b, p\in\mathcal{B}_{t,d}, n\in\mathcal{B}\setminus\mathcal{B}_{t,d}\}$, $\mathcal{D}_b$ is the bottom description set, $\mathcal{B}_{t,d}$ is the positive bottom set for the pair $(\textbf{I}_t,\textbf{d})$ and $\mathcal{B}\setminus\mathcal{B}_{t,d}$ is the negative bottom set for the pair $(\textbf{I}_t,\textbf{d})$.
The whole framework can be efficiently trained using back-propagation in an end-to-end paradigm.

For top recommendation, we follow the same way to build and train the model, but exchange the roles of tops and bottoms.

% !TEX root = ./www2019-fp-yujie-pengjie.tex

\section{Experimental Setup}
We set up a series of experiments to evaluate the recommendation performance of \ac{FARM}.
Details of our experimental settings are listed below.
All code and data used to run the experiments in this paper are available at \url{https://bitbucket.org/Jay_Ren/www2019_fas}\\\url{hionrecommendation_yujie/src/master/farm/}.

\subsection{Datasets}

Existing fashion datasets include \textit{WoW}~\cite{Liu2012}, \textit{Exact Street2Shop}~\cite{7410739}, \textit{Fashion-136K}~\cite{Jagadeesh2014}, \textit{FashionVC}~\cite{Song2017} and \textit{ExpFashion}~\cite{2018arXiv180608977L}.
\textit{WoW}, \textit{Exact Street2Shop}, and \textit{Fashion-136K} have been collected from street photos\footnote{\url{http://www.tamaraberg.com/street2shop/}} on the web and involve (visual) parsing of clothing, which still remains a great challenge in the computer vision domain \cite{Yamaguchi2012,6888484,Song2017} and which is beyond the scope of this paper.
\textit{FashionVC} and \textit{ExpFashion} have been collected from the fashion-oriented online community Polyvore\footnote{\url{http://www.polyvore.com/}} and contain both images and texts.
The images are of good quality and the texts include descriptions like names and categories.
For our experiments, we choose \textit{FashionVC} and \textit{ExpFashion}.
The statistics of the two datasets are given in Table~\ref{dataset}.
\begin{table}[h]
\small
\centering
\caption{Dataset statistics.}
\label{dataset}
\begin{tabular}{lrrr}
\toprule
\bf Dataset	& \bf Tops	& \bf Bottoms	& \bf Outfits\\
\midrule
FashionVC~\cite{Song2017}	& 14,871 & 13,663  & 20,726\\
ExpFashion~\cite{2018arXiv180608977L}	& 168,682 & 117,668  & 853,991\\
\bottomrule
\end{tabular}
\end{table}
We preprocess \textit{FashionVC} or \textit{ExpFashion} with the following steps, taking
bottom recommendation as an example.
For each tuple $(\mathit{top}, \mathit{top}\ \mathit{description}, \mathit{bottom}, \mathit{bottom}\ \mathit{description})$, we regard $(\mathit{top},$ $\mathit{bottom}\ \mathit{description})$ as input and the \emph{bottom} as the ground truth output.
We follow existing studies~\cite{Song2017} and randomly select bottoms to generate 100 candidates along with the ground truth bottoms in the validation and test set.
Similar processing steps are used for top recommendation.

\subsection{Implementation details}
The parameters $W$, $H$, $D$ and $N$ of the encoder and the generator are set to 1, 1, 1024 and 1024, respectively.
The size $e$ of the visual semantic word embedding, the semantic representation and the visual representation is set to 100.
And the latent variable size $k$ is set to 100 too.
The 7th, the 6th and the 5th layers of the encoder \ac{CNN} are adopted to compute the layer-to-layer matching with the input, the 1st and the 2nd layers of the generator \ac{DCNN}.
To build descriptions, we first filter out words whose frequency is less than 100.
Then, we manually go through the rest to only keep words that can describe tops or bottoms.
Finally, the remaining vocabulary size $D_d$ is 547.
During training, we initialize model parameters randomly with the Xavier method~\citep{Glorot2010Understanding}.
We choose Adam~\citep{Kingma2014Adam} as our optimization algorithm.
For the hyper-parameters of the Adam optimizer, we set the learning rate $\alpha$ = 0.001, two momentum parameters $\beta$1 = 0.9 and $\beta$2 = 0.999, and $\epsilon$ = $10^{-8}$.
We apply dropout \cite{Srivastava2014Dropout} to the output of our encoder and set the rate to 0.5.
We also apply gradient clipping \cite{Pascanu2013} with range $[-5, 5]$ during training.
We use a mini-batch size 64 by grid search to both speed up the training and converge quickly.
We test the model performance on the validation set for every epoch.
Our framework is implemented with MXNet~\cite{Chen2015MXNet}.
All experiments are conducted on a single Titan X GPU.

\subsection{Methods used for comparison}
We choose the following methods for comparison.

\begin{itemize}[nosep,leftmargin=*]
\item \textit{LR}: Logistic Regression (LR) is a standard machine learning meth- od~\cite{James2013An}.
We use it to predict whether a candidate bottom matches a given $(\mathit{top},\mathit{bottom}\ \mathit{description})$ pair or not.
Specifically, we employ a pre-trained \ac{CNN} to extract visual features from images.
Then we follow Eq.~\ref{LR} to calculate the matching probability $p$:
\begin{equation}
\label{LR}
p = \mathrm{sigmoid}(\textbf{w}_t^T\textbf{v}_t+\textbf{w}_b^T\textbf{v}_b+\textbf{w}_d^T\textbf{d}),
\end{equation}
where $\textbf{v}_t$ and $\textbf{v}_b\in\mathbb{R}^{D_v}$ are the visual features of the top and the bottom respectively, $\textbf{w}_t$ and $\textbf{w}_b\in\mathbb{R}^{D_v}$, and $\textbf{w}_d\in\mathbb{R}^{D_d}$.
$D_v$ is set to 4096 in our experiments.

\item \textit{IBR}$_d$: IBR~\cite{McAuley2015} learns a visual style space in which related objects are close and unrelated objects are far.
In order to consider the given descriptions at the same time, we modify IBR by projecting descriptions to the visual style space.
As a result, we can evaluate the matching degree between objects and descriptions by their distance in the space.
Specifically, the distance function between the candidate bottom $b$ and the given $(\mathit{top},\mathit{bottom}\ \mathit{description})$ pair $(t,d)$ is as follows:
\begin{equation}
m_{tdb} = \|\textbf{W}_v\textbf{v}_t-\textbf{W}_v\textbf{v}_b\|^2_2+\|\textbf{W}_d\textbf{v}_d-\textbf{W}_v\textbf{d}\|^2_2,
\end{equation}
where $\textbf{W}_v\in\mathbb{R}^{K\times{D_v}}$, $\textbf{W}_d\in\mathbb{R}^{K\times{D_d}}$, $\textbf{v}_t$ and $\textbf{v}_b\in\mathbb{R}^{D_v}$ are the visual features extracted by a pre-trained \ac{CNN}, and $K$ is the dimension of the visual style space.
$D_v$ is 4096, and $K$ is 100 in our experiments.
We refer to the modified version as IBR$_d$.

\item \textit{BPR-DAE}$_d$: BPR-DAE~\cite{Song2017} can jointly model the implicit matching preference between items in visual and textual modalities and the coherence relation between different modalities of items.
In our task, we do not have other text information except descriptions, so we first remove the part of BPR-DAE that is related to text information.
Then, for evaluating the matching score between the given description and the candidate item, we project the description representation and the item representation to the same latent space:
\begin{equation}
\textbf{v}_d' = \mathrm{sigmoid}(\textbf{W}_d\textbf{d}),\quad
\textbf{v}_i' = \mathrm{sigmoid}(\textbf{W}_v\textbf{v}_i),
\end{equation}
where $\textbf{W}_d\in\mathbb{R}^{K\times{D_d}}$, $\textbf{W}_v\in\mathbb{R}^{K\times{D_v}}$, and $\textbf{v}_i\in\mathbb{R}^{D_v}$ is the latent representation of item $i$ learned by BPR-DAE.
Finally, we follow Eq.~\ref{bpr-dae} to evaluate the compatibility between a candidate bottom $b$ and a given $(\mathit{top},\mathit{bottom}\ \mathit{description})$ pair $(t,d)$:
\begin{equation}
\label{bpr-dae}
m_{tdb} = \textbf{v}_t^T\textbf{v}_b+\textbf{v}_d'^T\textbf{v}_b'.
\end{equation}
We set $D_v = 512$, and $K = 100$ in experiments.
We refer to the modified version as BPR-DAE$_d$.

\item \textit{DVBPR}$_d$: DVBPR~\cite{DBLP:conf/icdm/KangFWM17} learns the image representations and trains the recommender system jointly to recommend fashion items for users.
We adopt DVBPR to our task and refer to it as DVBPR$_d$.
Specifically, we first follow DVBPR to use a \ac{CNN}-F to learn image representations of tops and bottoms.
Then we calculate the matching score between a bottom and the given $(\mathit{top}, \mathit{bottom}$ $\mathit{description})$ pair by Eq.~\ref{dvbpr}:
\begin{equation}
\label{dvbpr}
m_{tdb} = \textbf{v}_t^T\textbf{v}_b+\textbf{v}_d^T\textbf{v}_b,
\end{equation}
where $\textbf{v}_t$ and $\textbf{v}_b\in\mathbb{R}^K$ are the image representations of the top and bottom respectively, $\textbf{v}_d\in\mathbb{R}^K$ is the description representation learned in the same way as \ac{FARM}, and $K$ is set to 100 in experiments.
\end{itemize}

\subsection{Evaluation metrics}
We employ \acfi{MRR} and \acfi{AUC} to evaluate the recommendation performance, which are widely used in recommender systems~\cite{Rendle2010,Zhang2013,DBLP:conf/cikm/LiRCRLM17}.

In the case of bottom recommendations, for example, MRR and AUC are calculated as follows:
\begin{equation}
\text{MRR}=\frac{1}{|\mathcal{Q}^{td}|}\sum_{i=1}^{|\mathcal{Q}^{td}|} \frac{1}{\mathit{rank}_i},
\end{equation}
where $\mathcal{Q}^{td}$ is the $(\mathit{top}, \mathit{bottom}\ \mathit{description})$ collection as queries, and $\mathit{rank}_i$ refers to the rank position of the first positive bottom for the $i$-th $(\mathit{top},\mathit{bottom}\ \mathit{description})$ pair.
Furthermore,
\begin{equation}
\text{AUC}=\frac{1}{|\mathcal{Q}^{td}|} \sum_{(t,d)\in{\mathcal{Q}^{td}}} \frac{1}{|E(t,d)|}\sum_{(p,n)\in{E(t,d)}}\delta(s_p > s_n),
\end{equation}
where $E(t,d)$ is the set of all positive and negative candidate bottoms for the given top $t$ and the given bottom description $d$, $s_p$ is the matching score of a positive bottom $p$, $s_n$ is the matching score of a negative bottom $n$, and $\delta(\alpha)$ is an indicator function that equals 1 if $\alpha$ is true and 0 otherwise.

% !TEX root = ./www2019-fp-yujie-pengjie.tex

\section{Results}
\begin{table}[h]
\small
\centering
\caption{Recommendation results on the FashionVC and ExpFashion datasets (\%).}
\label{rec_results}
\begin{tabular}{l cc cc}
\toprule
\multirow{4}{*}{Method} & \multicolumn{4}{c}{\bf FashionVC} \\
\cmidrule(r){2-5}
& \multicolumn{2}{c}{Top} & \multicolumn{2}{c}{Bottom} \\
\cmidrule(r){2-3} \cmidrule(r){4-5}
& \multicolumn{1}{c}{AUC} & \multicolumn{1}{c}{MRR} & \multicolumn{1}{c}{AUC} & \multicolumn{1}{c}{MRR} \\
\midrule
LR & 48.7 & \phantom{1}4.5 & 46.4 & \phantom{1}4.4 \\
IBR$_d$ & 52.8 & \phantom{1}6.1 & 62.9 & 10.3 \\
BPR-DAE$_d$ & 62.9 & \phantom{1}8.6 & 70.2 & 10.9 \\
DVBPR$_d$ & 64.6 & \phantom{1}9.1 & 76.9 & 13.0 \\
\ac{FARM} & \textbf{71.2}\smash{\rlap{$^*$}} & \textbf{12.6}\smash{\rlap{$^*$}} & \textbf{77.8} & \textbf{15.3}\smash{\rlap{$^*$}} \\
\midrule
%\bottomrule
%\end{tabular}
%\begin{tabular}{l cc cc}
%\toprule
\multirow{4}{*}{Method} & \multicolumn{4}{c}{\bf ExpFashion} \\
\cmidrule(r){2-5}
& \multicolumn{2}{c}{Top} & \multicolumn{2}{c}{Bottom} \\
\cmidrule(r){2-3} \cmidrule(r){4-5}
& \multicolumn{1}{c}{AUC} & \multicolumn{1}{c}{MRR} & \multicolumn{1}{c}{AUC} & \multicolumn{1}{c}{MRR} \\
\midrule
LR & 50.5 & \phantom{1}5.4 & 48.4 & \phantom{1}4.4 \\
IBR$_d$ & 56.1 & \phantom{1}7.1 & 68.9 & 12.0 \\
BPR-DAE$_d$ & 73.0 & 12.3 & 79.9 & 14.7 \\
DVBPR$_d$ & 82.4 & 18.5 & 83.7 & 15.4 \\
\ac{FARM} & \textbf{85.2}\smash{\rlap{$^*$}} & \textbf{25.1}\smash{\rlap{$^*$}} & \textbf{88.4}\smash{\rlap{$^*$}} & \textbf{24.3}\smash{\rlap{$^*$}} \\
\bottomrule
\end{tabular}
\begin{minipage}{\columnwidth}
The superscript $^*$ indicates that \ac{FARM} significantly outperforms DVBPR$_d$, using a paired t-test with $p<0.05$.
\end{minipage}
\end{table}

The recommendation results on the FashionVC and ExpFashion datasets of \ac{FARM} and the methods used for comparison are shown in Table~\ref{rec_results}.
We can see that \ac{FARM} consistently outperforms all baselines in terms of AUC and MRR on both datasets.
We have five main observations from Table~\ref{rec_results}.
\begin{enumerate}[leftmargin=*,nosep]
\item \ac{FARM} significantly outperforms all baselines and achieves the best results on all metrics.
There are three main reasons.
First, \ac{FARM} contains a fashion generator as an auxiliary module for recommendation.
With its co-supervision learning framework, \ac{FARM} can encode more aesthetic characteristics and use this extra information to improve recommendation performance; see Section~\ref{section:mla} for further analysis.
Second, we propose a layer-to-layer matching scheme to make sure that \ac{FARM} can effectively use the aesthetic features in the fashion generator to improve recommendation results; see Section~\ref{section:pma} for a further analysis.
Third, LR, IBR$_d$ and BPR-DAE$_d$ employ pre-trained \acp{CNN} (all AlexNet~\cite{Jia2014} trained on ImageNet\footnote{\url{http://www.image-net.org/}}) to extract visual features from images, but they do not fine-tune the \acp{CNN} during experiments.
However, in \ac{FARM}, we jointly train the top encoder, the bottom encoder and the top/bottom generator, which can extract better visual features.

\item DVBPR$_d$ performs better than other baseline methods.
The reason is that DVBPR$_d$ employs a CNN-F to jointly learn image representations during recommendation.
Hence, it can extract more effective visual features to improve recommendation performance.

\item Although BPR-DAE$_d$, IBR$_d$ and LR all use visual features extracted by a pre-trained \ac{CNN} as input, BPR-DAE$_d$ performs much better than the other two.
This is because BPR-DAE$_d$ learns a more sophisticated latent space using an auto-encoder neural network to represent the fashion items.
However, IBR$_d$ only applies a linear transformation to inputs, which restricts the expressive ability of the visual style space.
And LR directly uses the visual features and the bag-of-words vectors as inputs, making it hard to learn an effective matching relation.

\item The performance of all methods on the ExpFashion dataset is better than on the FashionVC dataset.
The most important reason is that the average length of the descriptions in the ExpFashion dataset is 5.6 words, however, it is only 3.7 words in the FashionVC dataset.
That means that the descriptions in the ExpFashion dataset contain more details that can provide more information for recommendation and generation, which boosts the recommendation performance.

\item The bottom recommendation performance is better than the top recommendation performance for most methods.
The number of tops is larger than the number of bottoms and the styles of tops are also richer than those of bottoms on both datasets.
That makes bottom recommendation and bottom generation easier.
\end{enumerate}

\noindent%
% \todo{Include a final upshot for this section.}
In summary, \ac{FARM} significantly outperforms state-of-the-art methods on both datasets. The improvements mainly come from the co-supervision of generation and the layer-to-layer mechanism, which we will demonstrate in the next section.

% !TEX root = ./www2019-fp-yujie-pengjie.tex

\section{Analysis}

We provide two types of analyses (concerning co-supervision learning and layer-to-layer matching) and two cases studies (recommendation and generation).

\subsection{Co-supervision learning}
\label{section:mla}
\begin{table}
\small
\centering
\caption{Analysis of co-supervision learning. Recommendation results on the FashionVC and ExpFashion datasets (\%).}
\label{mul_ana_results}
\begin{tabular}{l cc cc}
\toprule
\multirow{4}{*}{Method} & \multicolumn{4}{c}{\bf FashionVC} \\
\cmidrule(r){2-5}
& \multicolumn{2}{c}{Top} & \multicolumn{2}{c}{Bottom} \\
\cmidrule(r){2-3} \cmidrule(r){4-5}
& \multicolumn{1}{c}{AUC} & \multicolumn{1}{c}{MRR} & \multicolumn{1}{c}{AUC} & \multicolumn{1}{c}{MRR} \\
\midrule
\ac{FARM}-G & 54.8 & \phantom{1}8.4 & 60.9 & \phantom{1}9.8 \\
\ac{FARM}-R & 68.0 & \phantom{1}9.8 & 77.2 & 12.8 \\
\ac{FARM} & \textbf{71.2}\smash{\rlap{$^*$}} & \textbf{12.6}\smash{\rlap{$^*$}} & \textbf{77.8} & \textbf{15.3}\smash{\rlap{$^*$}} \\
\midrule
%\bottomrule
%\end{tabular}
%\begin{tabular}{l cc cc}
%\toprule
\multirow{4}{*}{Method} & \multicolumn{4}{c}{\bf ExpFashion} \\
\cmidrule(r){2-5}
& \multicolumn{2}{c}{Top} & \multicolumn{2}{c}{Bottom} \\
\cmidrule(r){2-3} \cmidrule(r){4-5}
& \multicolumn{1}{c}{AUC} & \multicolumn{1}{c}{MRR} & \multicolumn{1}{c}{AUC} & \multicolumn{1}{c}{MRR} \\
\midrule
\ac{FARM}-G & 64.4 & 14.2 & 72.4 & 21.3 \\
\ac{FARM}-R & 82.3 & 18.9 & 84.2 & 15.2 \\
\ac{FARM} & \textbf{85.2}\smash{\rlap{$^*$}} & \textbf{25.1}\smash{\rlap{$^*$}} & \textbf{88.4}\smash{\rlap{$^*$}} & \textbf{24.3}\smash{\rlap{$^*$}} \\
\bottomrule
\end{tabular}
\begin{minipage}{\columnwidth}
The superscript $^*$ indicates that \ac{FARM} significantly outperforms \ac{FARM}-R, using a paired t-test with $p<0.05$.
\end{minipage}
\end{table}

To demonstrate the superiority of incorporating the extra supervision of the generator, we compare \ac{FARM} with \ac{FARM}-G and \ac{FARM}-R, where \ac{FARM}-G is \ac{FARM} without the recommendation part and \ac{FARM}-R is \ac{FARM} without the generation part.
The results are shown in Table~\ref{mul_ana_results}.
To be able to apply \ac{FARM}-G to the recommendation task, we first use \ac{FARM}-G to generate a bottom image for a given $(\mathit{top}, \mathit{bottom}\ \mathit{description})$ pair.
Then, similar to~\cite{Babenko2014NeuralCF,Lin2015Deep}, we use a pre-trained AlexNet to get the representations of the generated bottom and the candidate bottoms.
Finally, we compute the similarity between the generated bottom and a candidate bottom based on their representations.

From Table~\ref{mul_ana_results}, we can see that \ac{FARM} achieves significant improvements over \ac{FARM}-R.
On the FashionVC dataset, for top recommendation, AUC increases by 3.2\%, MRR increases by 2.8\%, and for bottom recommendation, AUC increases by 0.6\%, MRR increases by 2.5\%.
On the ExpFashion dataset, for top recommendation, AUC increases by 2.9\%, MRR increases by 6.2\%, and for bottom recommendation, AUC increases by 4.2\%, MRR increases by 9.1\%.
Thus, \ac{FARM} is able to improve recommendation performance by using the generator as a supervision signal.

Comparing \ac{FARM}-G with all baselines, we notice that \ac{FARM}-G achieves better performance, and especially it performs better than IBR$_d$ in most settings.
Hence, the images generated by \ac{FARM}-G and \ac{FARM} reflect some key factors of the items to be recommended, which is why the generator can help improve recommendation.

Additionally, we find that \ac{FARM}-R outperforms LR, IBR$_d$ and BPR-DAE$_d$.
And it achieves comparable performance with DVBPR$_d$, whose difference against \ac{FARM}-R is mainly in the \ac{CNN} part.
If \ac{FARM} employs more powerful \ac{CNN} architectures such as VGG~\cite{Simonyan2014Very} or ResNet~\cite{He2016Deep}, it should perform even better.

\subsection{Layer-to-layer matching}
\label{section:pma}

\begin{table}
\small
\centering
\caption{Analysis of layer-to-layer matching. Recommendation results on the FashionVC and ExpFashion datasets (\%).}
\label{p2p_ana_results}
\begin{tabular}{l cc cc}
\toprule
\multirow{4}{*}{Method} & \multicolumn{4}{c}{\bf FashionVC} \\
\cmidrule(r){2-5}
& \multicolumn{2}{c}{Top} & \multicolumn{2}{c}{Bottom} \\
\cmidrule(r){2-3} \cmidrule(r){4-5}
& \multicolumn{1}{c}{AUC} & \multicolumn{1}{c}{MRR} & \multicolumn{1}{c}{AUC} & \multicolumn{1}{c}{MRR} \\
\midrule
\ac{FARM}-WL & 59.8 & \phantom{1}7.6 & 67.8 & \phantom{1}8.2 \\
\ac{FARM} & \textbf{71.2}\smash{\rlap{$^*$}} & \textbf{12.6}\smash{\rlap{$^*$}} & \textbf{77.8}\smash{\rlap{$^*$}} & \textbf{15.3}\smash{\rlap{$^*$}} \\
\midrule
%\bottomrule
%\end{tabular}
%\begin{tabular}{l cc cc}
%\toprule
\multirow{4}{*}{Method} & \multicolumn{4}{c}{\bf ExpFashion} \\
\cmidrule(r){2-5}
& \multicolumn{2}{c}{Top} & \multicolumn{2}{c}{Bottom} \\
\cmidrule(r){2-3} \cmidrule(r){4-5}
& \multicolumn{1}{c}{AUC} & \multicolumn{1}{c}{MRR} & \multicolumn{1}{c}{AUC} & \multicolumn{1}{c}{MRR} \\
\midrule
\ac{FARM}-WL & 68.6 & \phantom{1}9.9 & 74.3 & 10.3 \\
\ac{FARM} & \textbf{85.2}\smash{\rlap{$^*$}} & \textbf{25.1}\smash{\rlap{$^*$}} & \textbf{88.4}\smash{\rlap{$^*$}} & \textbf{24.3}\smash{\rlap{$^*$}} \\
\bottomrule
\end{tabular}
\begin{minipage}{\columnwidth}
The superscript $^*$ indicates that \ac{FARM} significantly outperforms \ac{FARM}-WL, using a paired t-test with $p<0.05$.
\end{minipage}
\end{table}

To analyze the effect of the layer-to-layer matching scheme, we compare \ac{FARM} with \ac{FARM}-WL which only uses the visual matching and the description matching to evaluate the matching degree.
We can see from Table~\ref{p2p_ana_results} that \ac{FARM} performs significantly better than \ac{FARM}-WL according to all metrics on both datasets, which confirms that layer-to-layer matching does indeed improve the performance of recommendation.

\begin{figure}[h]
 \centering
 \subfigure[Top generation.]{
 \label{f_6_1}
 \includegraphics[clip,trim=0mm 0mm 0mm 0mm,width=1\columnwidth]{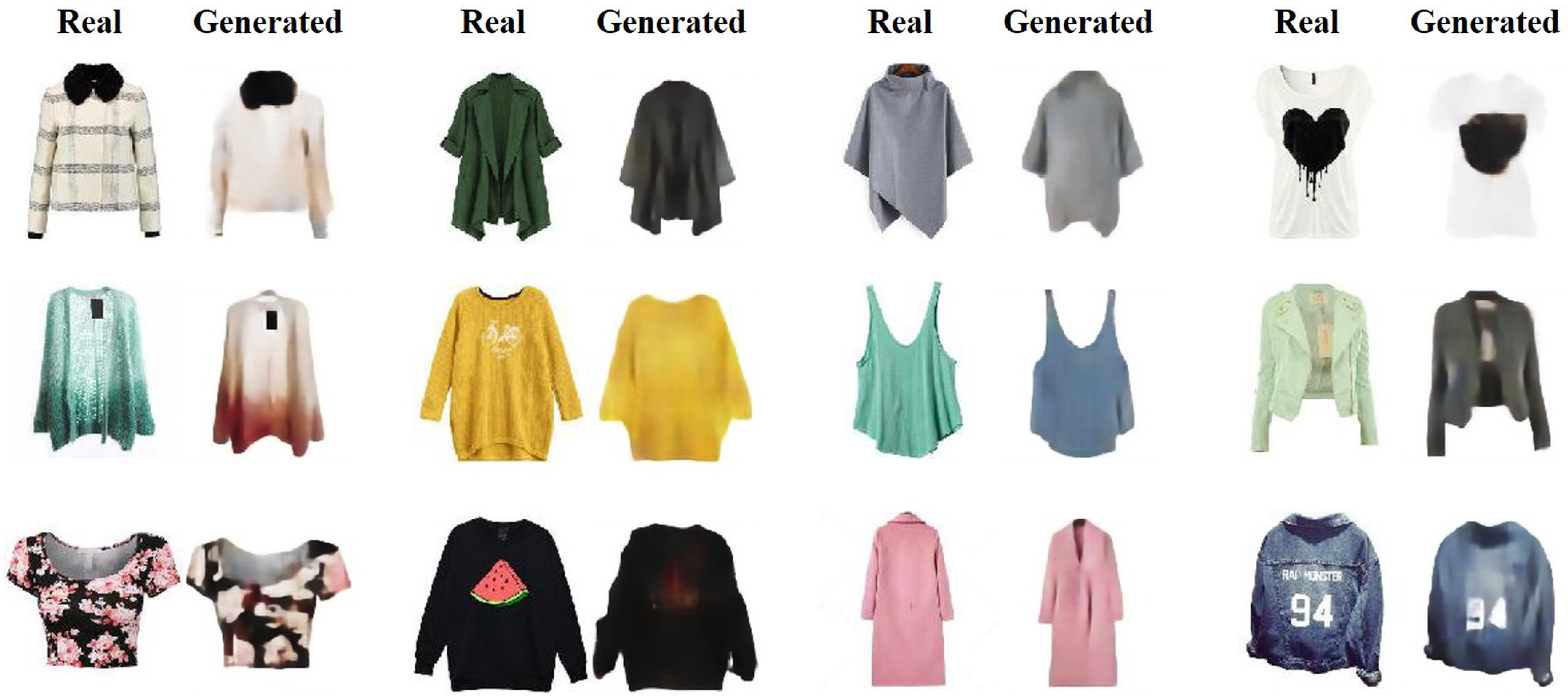}}
 \subfigure[Bottom generation.]{
 \label{f_6_2}
 \includegraphics[clip,trim=0mm 0mm 0mm 0mm,width=1\columnwidth]{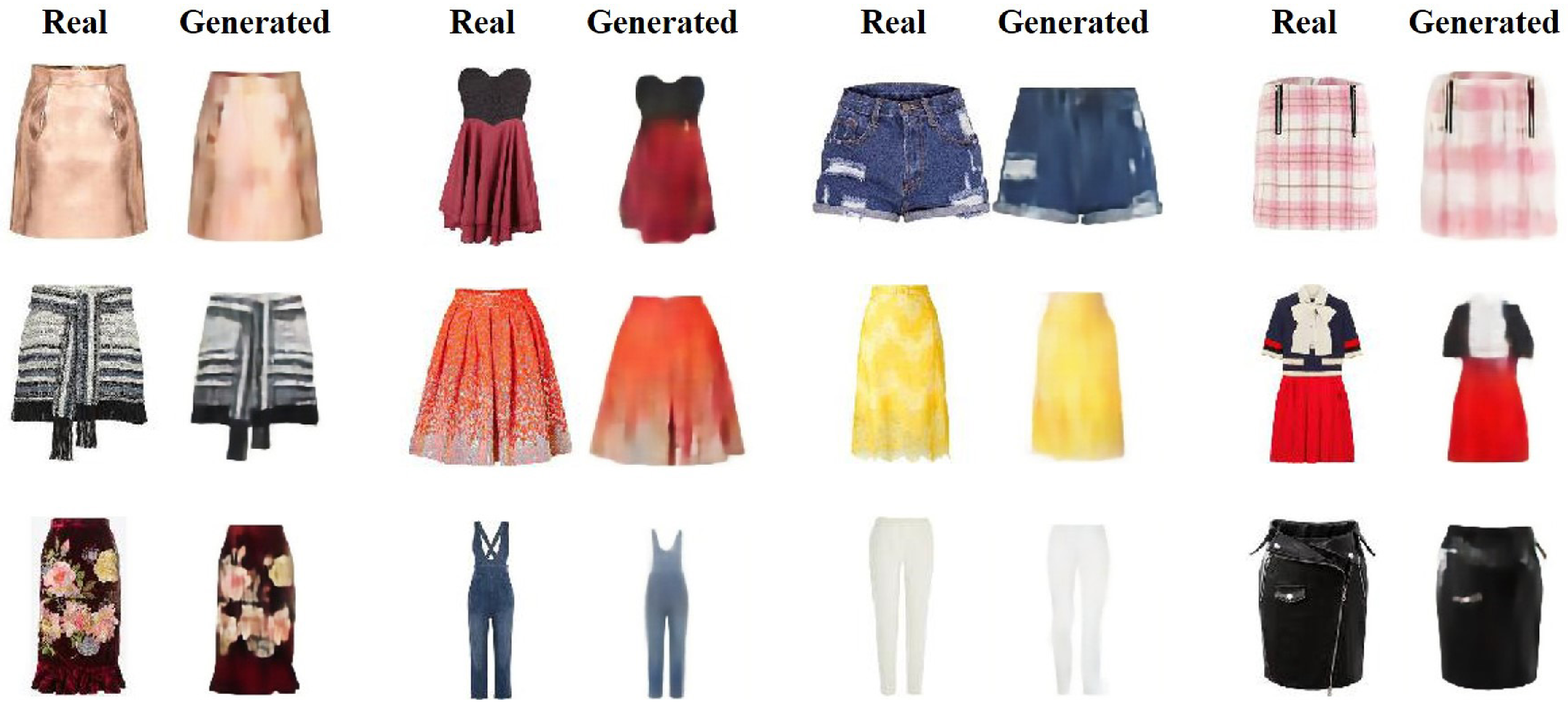}}
 \caption{Comparison between real and generated images.}
 \label{ana_figure_1}
\end{figure}

To help understand the effect of layer-to-layer matching, we list some real and generated images in Figure~\ref{ana_figure_1}.
\ac{FARM} generates good quality images that are similar to real images.
This means that the generated images can tell us what kind of bottoms can match the given $(\mathit{top},\mathit{bottom}\ \mathit{description})$ pair from the perspective of generation, so layer-to-layer matching can direct the recommender by evaluating the matching degree between the candidate images and the generated images.
That is why layer-to-layer matching is able to improve the performance of recommendation.

Additionally, we notice that \ac{FARM}-WL performs worse than \ac{FARM}-R, which means that a simple combination of recommendation and generation is not able to improve recommendation performance significantly.
This may be because, without layer-to-layer matching, \ac{FARM}-WL pays too much attention to the generation quality and ignores recommendation performance.
We are able to improve this situation with layer-to-layer matching.
Layer-to-layer matching builds a connection between the bottom generator and the bottom encoder in different layers.
As a result, the bottom encoder pushes the bottom generator to learn useful matching information for improving recommendation performance.

\subsection{Recommendation case studies}

\begin{figure}[h]
 \centering
 \subfigure[Top recommendation.]{
 \label{f_6_3}
 \includegraphics[clip,trim=4mm 0mm 0mm 0mm,width=1\columnwidth]{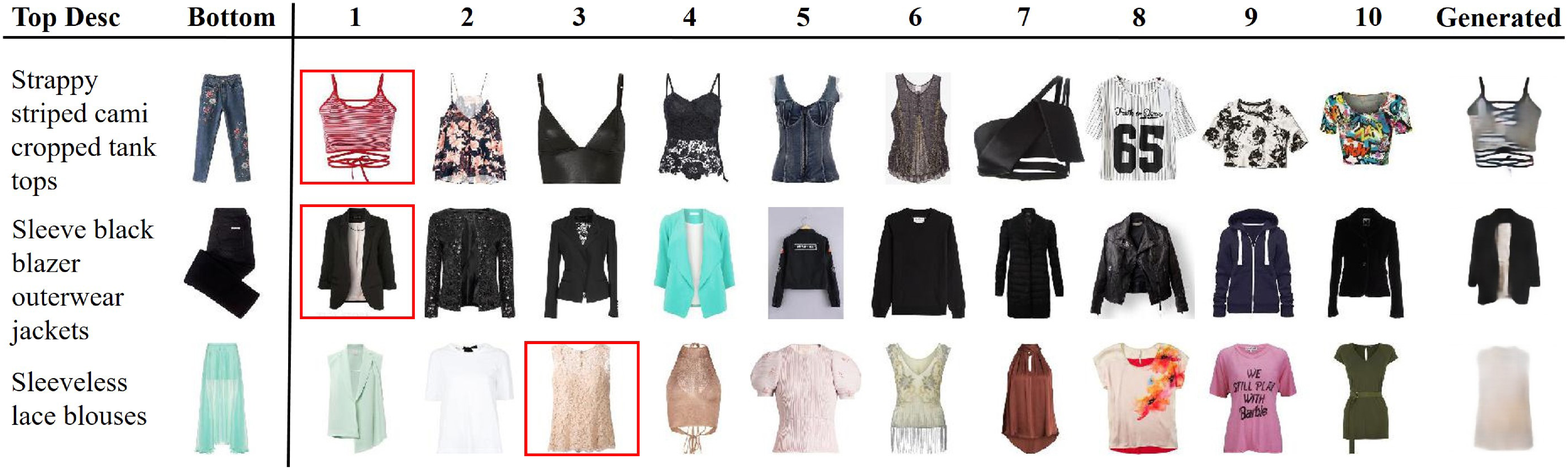}}
 \subfigure[Bottom recommendation.]{
 \label{f_6_4}
 \includegraphics[clip,trim=4mm 0mm 0mm 0mm,width=1\columnwidth]{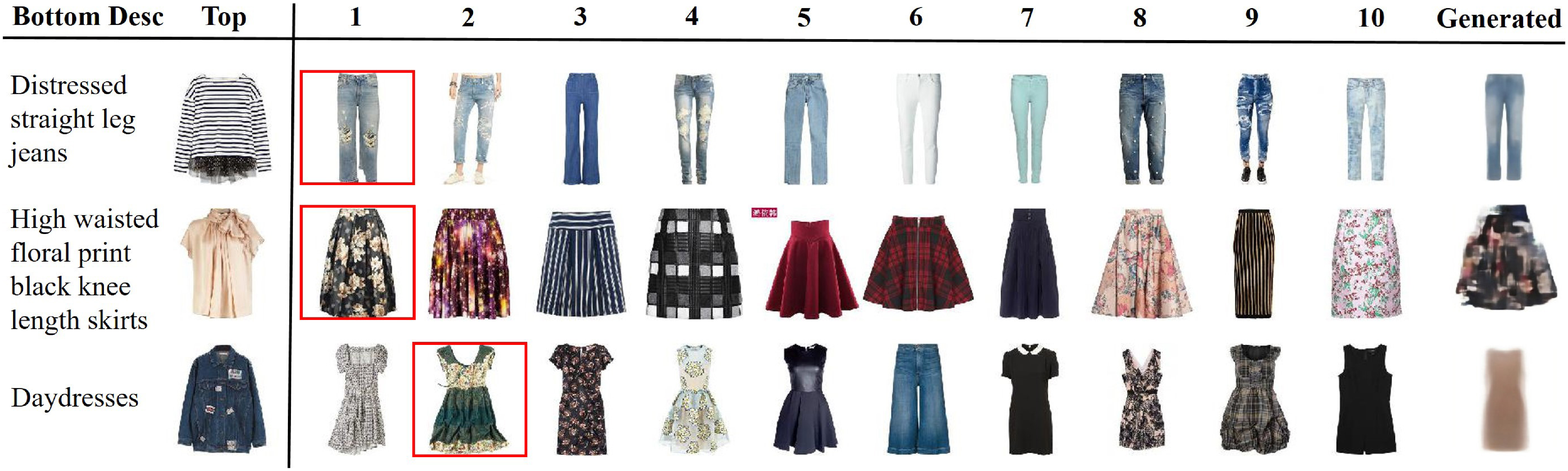}}
 \caption{Case studies of recommendation. The items highlighted in the red boxes are the positive ones.}
 \label{ana_figure_2}
\end{figure}

We list some recommendation produced by \ac{FARM} in Figure~\ref{ana_figure_2}.
For each input, we list the top-10 recommended items.
We highlight the positive items with red boxes.
We can see that most recommended items not only match the given items, but also meet the given descriptions.
For example, in the second case of the top recommendation, the given top description is ``sleeve black blazer outerwear jackets,'' so most recommended tops are jackets, and especially almost all recommended tops are black.
Also in the first case of the bottom recommendation, the given bottom description is ``distressed straight leg jean,'' so the recommended bottoms are all jeans, most of which are straight leg and some are distressed.
By comparing the generated items with the recommended items, such as in the first case of the top recommendation and the second case of the bottom recommendation, we can see that the generated images are able to provide good guidance for the recommendation.

We also notice that not all recommended items meet the given description, mostly because \ac{FARM} recommends items not only based on the given description, but also based on the given item.
For example, in the third case of the bottom recommendation, the sixth recommended bottom is a denim jeans instead of a daydress.
The given top is a denim coat, which makes \ac{FARM} believe that recommending a denim jeans is also reasonable.
Besides, not all positive items are ranked in the first position.
See, e.g., the third case of the top recommendation., where the top recommended item and the given bottom have the same color green, which looks more compatible.
In these failure cases, the quality of the generated images is poor so they are likely less helpful for recommendation.

\subsection{Generation case studies}

\begin{figure}[htb]
 \centering
 \subfigure[Top generation.]{
 \label{f_6_5}
 \includegraphics[clip,trim=5mm 0mm 0mm 0mm,width=1\columnwidth]{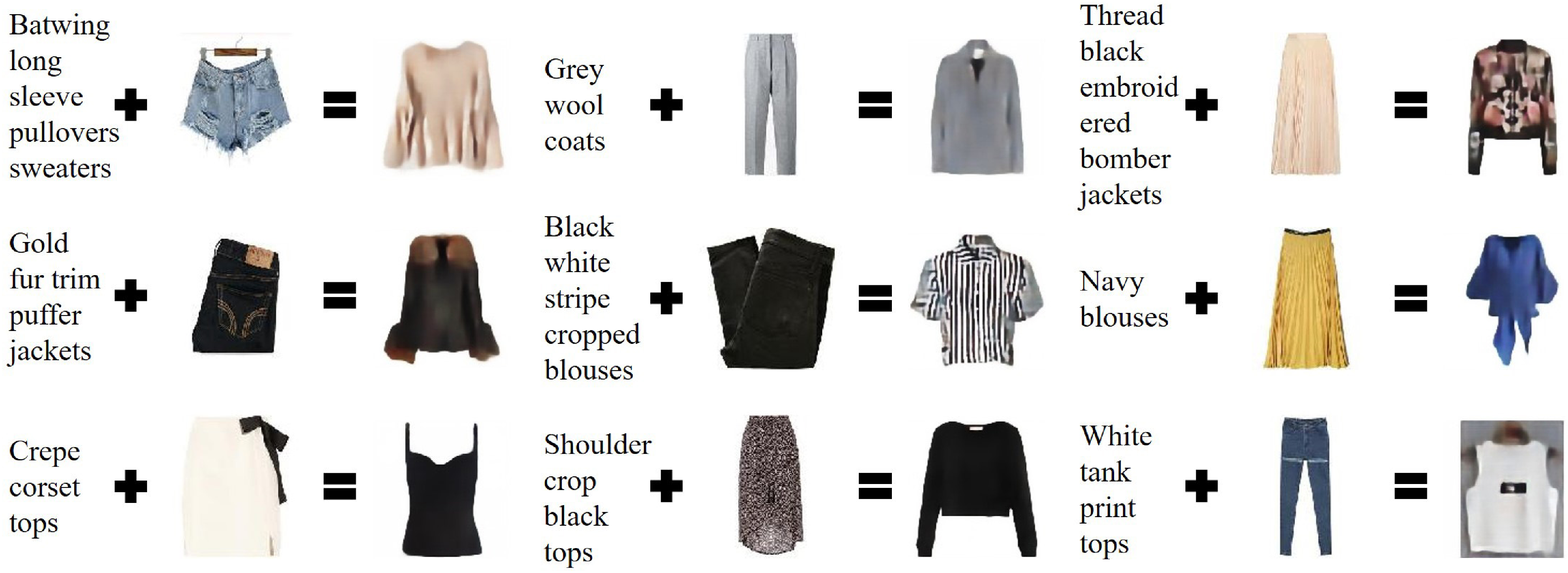}}
 \subfigure[Bottom generation.]{
 \label{f_6_6}
 \includegraphics[clip,trim=5mm 0mm 0mm 0mm,width=1\columnwidth]{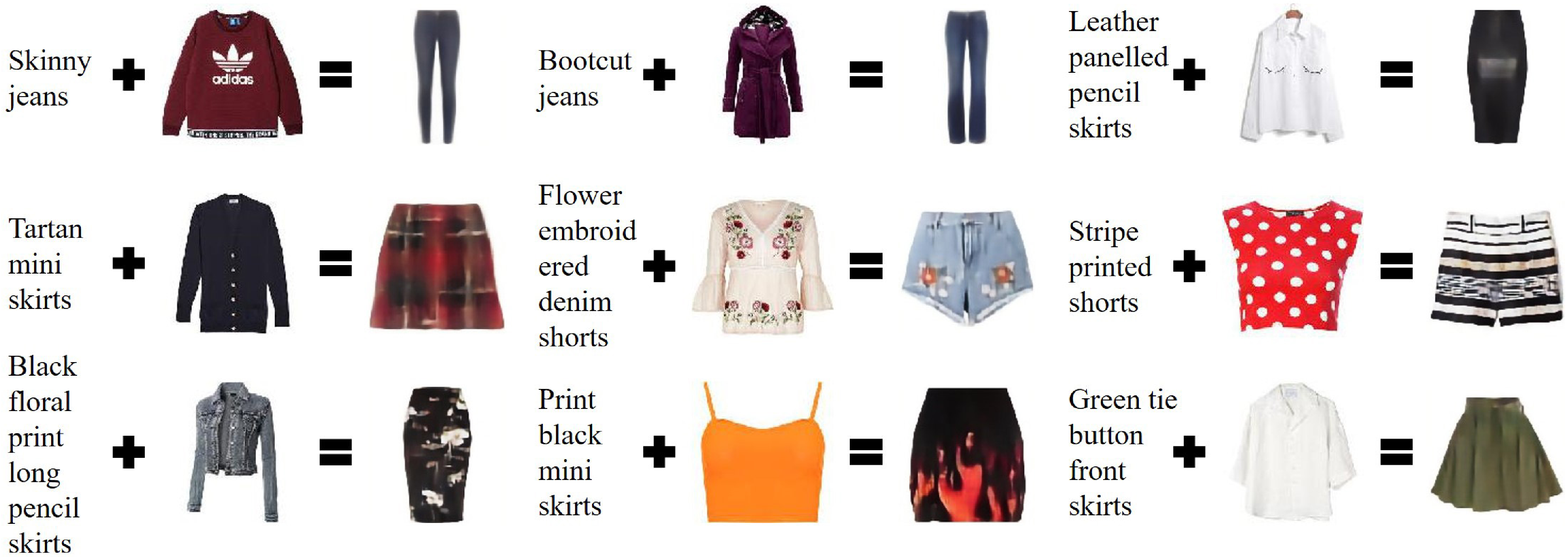}}
 \caption{Case studies of generation. Each case is in the form: ``given description + given item = generated item''.}
 \label{ana_figure_3}
\end{figure}

Although this paper focuses on improving recommendation by incorporating generation, we also list some generation cases in Figure~\ref{ana_figure_3}.
Overall, the generated items are able to match the given input.
For example, in the sixth case of the top generation, the generated navy blouse with the yellow keen length skirt looks beautiful and elegant.
Although there are many kinds of navy blouses like sailor suits, the style of the generated top seems to be more suitable for the given bottom.
And in the eighth case of the bottom generation, the given description does not give the specific pattern of the generated bottom.
But the generated bottom has a flame-like pattern, which makes it more compatible with the bright yellow camisole.
From these samples we can see that \ac{FARM} is able to generate fashion items based on the relation between the visual features of different fashion items.

The generated items can accord with the given descriptions no matter what they are about.
For example, in the second case of the top generation, the description is ``grey wool coats,'' so the generated top is a grey coat which also looks like wool.
And in the fourth case, the description is ``gold fur trim puffer jackets'', so the generated jacket has fur in its collar and cuff.
In the bottom generation, we also observe that \ac{FARM} is able to distinguish between skinny jeans and bootcut jeans from the first and the second cases.
Another example is the sixth case, where the description contains ``floral print.''
\ac{FARM} generates a black long pencil skirt with flower pattern.
In short, \ac{FARM} is able to build a cross-modal connection between text and images in order to generate fashion items.

Generation is a challenging process, which means that powerful features are needed in order to generate a matching item.
We can see from the examples provided that \ac{FARM} is able to generate aesthetically matching outfits.
\ac{FARM} is able to improve recommendation performance through jointly modeling generation.

% !TEX root = ./www2019-fp-yujie-pengjie.tex

\section{Conclusion}
In this paper, we have studied the task of outfit recommendation, which has two main challenges: \acl{VU} and \acl{VM}. 
To tackle these challenges, we propose a co-supervision learning framework, namely \ac{FARM}. For \acl{VU}, \ac{FARM} captures aesthetic characteristics with the supervision of generation learning.
For \acl{VM}, \ac{FARM} incorporates a layer-to-layer matching mechanism to evaluate the matching score at different neural layers.

We have conducted experiments to confirm the effectiveness of \ac{FARM}. 
It achieves significant improvements over state-of-the-art baselines in terms of \acs{AUC} and \acs{MRR}. 
We also show that the proposed layer-to-layer matching mechanism can make effective use of generation information to improve recommendation performance. 
We further exhibit some cases to analyze the  performance of \ac{FARM}.

Our results can be used to improve users' experience in fashion-oriented online communities by providing better recommendation and to promote the research into fashion generation by demonstrating a novel application in outfit recommendation.

A limitation of \ac{FARM} is that its recommendation performance is affected by the quality of the generated images. 
If the quality of the generated images is not high, the generation part cannot provide effective guidance for the recommendation part.

As to future work, we plan to improve the recommendation and the generation of \ac{FARM} when the descriptions are lacking. 
And we want to extend \ac{FARM} to recommend and generate whole outfits that not only contain tops and bottoms but also include shoes and hats, etc. 
We will also try more powerful \ac{CNN} and \ac{DCNN} architectures for recommendation and generation.

\section*{Acknowledgments}
We thank the anonymous reviewers for their helpful comments.

This work is supported by the Natural Science Foundation of China (61672324, 61672322), the Natural Science Foundation of Shandong province (2016ZRE27468), the Fundamental Research Funds of Shandong University, Ahold Delhaize, the Association of Universities in the Netherlands, and the Innovation Center for Artificial Intelligence (ICAI).
All content represents the opinion of the authors, which is not necessarily shared or endorsed by their respective employers and/or sponsors.

\newpage

\bibliographystyle{ACM-Reference-Format}
\bibliography{www2019-fp-yujie-pengjie}

\end{document}